\documentclass[12pt]{article}
\usepackage{epsfig}
\begin{document}
\begin{center}
{\large \bf 
On the Ter-Mikaelian and Landau--Pomeranchuk effects for 
induced soft gluon radiation in a QCD medium}\\[6mm]
{\sc B. K\"ampfer$^1$, O.P. Pavlenko$^{1,2}$}\\[6mm]
$^1$Forschungszentrum Rossendorf, PF 510119, 01314 Dresden, Germany\\
$^2$Institute for Theoretical Physics, 252143 Kiev - 143, Ukraine
\end{center}

\centerline{Abstract}
The polarization of a surrounding QCD medium 
modifies the induced gluon radiation spectrum
of a high-energy parton at small transverse momenta
for a single interaction and for multiple scatterings as well.
This effect is an analogue of the Ter-Mikaelian effect in QED,
superimposed to the Landau--Pomeranchuk effect,
however it appears in QCD in a different phase space region.

\vspace*{3cm}

{\bf Introduction:}
The soft gluon radiation induced by energetic partons propagating through a
medium of quarks and gluons is of great interest now, since the radiative 
energy loss and corresponding stopping power of quark or gluon jets might 
serve as a probe of the quark-gluon plasma formation in ultrarelativistic 
heavy-ion collisions
(cf. \cite{Wang,GW} and references therein).
It was recently shown \cite{GW,Baier,Zakharov}
that the Landau--Pomeranchuk--Migdal (LPM) effect 
(cf. \cite{Migdal})
plays a very important role for the formation of 
gluon bremsstrahlung in a QCD medium. In analogy with QED the 
LPM effect for gluon radiation is related to a destructive interference
between radiation amplitudes due to multiple scatterings
of a high-energy color charge propagating through the medium.
The crucial physical characteristics here is the radiation formation
length $l_f$. To radiate coherently the emitting particle must be 
undisturbed while traversing the formation length. Multiple scatterings 
within $l_f$ can cause a loss of coherence. This leads to a suppression
and qualitative change of the soft radiation spectrum compared to the
well-known Bethe--Heitler formula, where the spectrum is simply
additive in the number of scatterings. It should be stressed that the LPM
effect becomes operative in a rather restricted kinematical region 
of the energy $E$ of the radiating particle and energy $\omega$
of the radiation: the formation length must be large in comparison
with the mean free path $\lambda$, $l_f \gg \lambda$. In particular, to fulfill
this condition in the case of an extended medium, modeled by screened
static scattering centers \cite{GW}, the relations 
$\omega \ll E^2 / (\lambda \mu^2)$ within QED or 
$\omega \gg \lambda \mu^2$ within QCD must be fulfilled \cite{Baier}
($\mu$ is the Debye screening mass).

At the same time it is well-known from electrodynamics that the formation
length can be modified not only by multiple scatterings of the radiating
particle but also by the medium polarization, i.e. the emitted
photon is affected. It was firstly pointed out by Ter-Mikaelian 
\cite{TM,TM_textbook}
that the dielectric polarization  of the medium can also cause a loss of
coherence, suppressing in this way the emission process.
This effect, also known as dielectrical effect or longitudinal density
effect, suppresses the very soft bremsstrahlung photons, while the LPM 
effect becomes operative at larger photon energies.

The strong reduction of the formation length, reflecting the suppression of
radiation due to the dielectrical effect,
can be seen in electrodynamics by the following
simple qualitative estimates. In the high-frequency approximation the
dielectrical ''constant'' becomes $\epsilon = 1 - \omega_p^2/ \omega^2$,
where $\omega_p$ is the plasma frequency. In a medium the dispersion relation
between energy $\omega$ and momentum $\vec k$ becomes 
$\omega = |\vec k | / \sqrt{\epsilon}$ in contrast to the vacuum
dispersion relation  $\omega = |\vec k |$
(we use units with $\hbar = c =1$).
As a result the ''vacuum'' formation length $l_f = 2 \gamma^2 / \omega$
is reduced to $l_f = 2 \omega / \omega_p^2$ in the interval
$\omega_p \ll \omega \ll \gamma \omega_p$, where $\gamma$ is the Lorentz
factor of the radiating particle.

The Ter-Mikaelian (TM) effect is important in electrodynamics 
because it cuts off the soft bremsstrahlung spectrum at low energies, 
thus removing the infrared divergence. 
The recent experiments with high-energy electrons at SLAC \cite{Anthony} 
confirmed the TM effect, predicted 45 years ago.

Due to the long-range properties of color forces one can expect that the
polarization of a QCD medium is also important for the induced gluon
radiation. A throughout consideration of this problem could be based on the
gluon polarization tensor in a hot medium 
\cite{Kalashnikov,Klimov}
together with proper modifications of the propagators and vertices
in the gluon radiation Feynman diagrams. However, even for the 
Abelian theory such a consideration appears rather involved \cite{Migdal}, 
and a simpler
approach provides better insight in the problem \cite{TM_textbook}.
Fortunately, in the high-temperature limit the needed dispersion relation
for gluons  in a QCD medium \cite{Klimov} can be approximated by introducing
an effective gluon mass $\omega_0$ \cite{Biro,Peshier,Levai}, 
which depends on
the temperature $T$ (we consider a charge-symmetric medium).
Within this approximation the dispersion relation for gluons has a form
similar to the Abelian case \cite{LL}
\begin{equation}
k^2 \equiv k^\mu k_\mu = \omega^2 - k_\parallel^2 - k_\perp^2 = \omega_0^2,
\label{eq.1}
\end{equation} 
where $k^\mu = (\omega, k_\parallel, k_\perp)$ is the gluon
four-momentum, and $\omega_0 (T)$ parameterizes the temperature dependent
gluon self-energy. Such an effective quasi-particle model of massive gluons
is proven successful in reproducing the results of lattice calculations
for various collective properties of the hot QCD medium, such as the
equation of state and Debye screening mass \cite{Peshier}.
Moreover, a detailed analysis \cite{Peshier,Levai} 
of the recent QCD lattice data \cite{Karsch}
allows to extract $\omega_0 (T)$.
Typically $\omega_0$ is in the order of a few hundred MeV already
slightly above the confinement temperature.

Here we are going to consider the influence of the 
QCD medium polarization on the induced soft gluon emission and demonstrate 
the existence of the  non-Abelian analogue to the TM effect.
Basing on the gluon dispersion relation (1) we begin treating the gluon
radiation for single scattering of a fast parton and show that, due to the
polarization of the surrounding medium, the resulting gluon spectrum 
in the small
transverse momentum region is considerably suppressed and the infrared
divergence at $k_\perp \to 0$ is regularized by the dimension
parameter $\omega_0$. 
Then this result is extended to multiple scatterings in a QCD medium
within the potential model \cite{GW}.
In the region of small transverse momenta of the radiated gluons, which is
relevant for the TM effect, we analyze  the medium corrections for this 
specific QCD radiation which is actually absent in the Abelian case.
Another influence of the medium on the radiation is the multiple gluon
scattering due to the non-Abelian interaction considered in ref.~\cite{Baier}.
Here we focus on single-gluon emission; multiple gluon emission is dealt with
in ref.~\cite{Gyulassy_Levai}.

%%%%%%%%%%%%%%%%%%%%%%%%%%%%%%%%%%%%%%%%%%%%%%%%%%%%%%%%%%%%%%%%%%%%%%%

{\bf Single scattering:}
To demonstrate the importance of the modified dispersion relation eq.~(1)
we consider first the induced gluon radiation for a single quark-quark
scattering. The cross section for this process was originally derived by 
Gunion and Bertsch \cite{GB}. Following their calculations we use the light
cone representation of four-vectors and the $A^+ = 0$ gauge for the
gluon field. 
The variables for the radiation process are defined in fig.~1. In the
center-of-mass frame of the colliding quarks the initial momenta of quarks and
and the final gluon and its polarization are
\begin{eqnarray}
p_i & = & [\sqrt{s},0,\vec 0_\perp],\quad
p_i' =  [0, \sqrt{s},\vec 0_\perp],\\
k & = & [x \sqrt{s}, (k_\perp^2 + \omega_0^2)/(x \sqrt{s}), \vec k_\perp],
\quad
\epsilon = [0, 2 \vec k_\perp \vec \epsilon_\perp / (x \sqrt{s}), 
\vec \epsilon_\perp],
\end{eqnarray} 
where $x = (\omega + k_\parallel) / \sqrt{s}$ and $s = (p_i + p_i')^2$.
The components of the gluon polarization obey the
condition $\epsilon k = 0$ which is valid for massive bosons too \cite{LL}.
Using the condition that the quarks after scattering are on-mass shell 
one derives for $x \ll 1$ the momentum transfer as
\begin{equation}
q =  [- \frac{q_\perp^2}{\sqrt{s}},
\frac{x^{-1} k_\perp^2 + q_\perp^2 - 2 \vec q_\perp \vec k_\perp}
{(1-x) \sqrt{s}} 
+ 
\frac{\omega_0^2}{x \sqrt{s}},
\vec q_\perp] .
\end{equation}
As shown in ref.~\cite{GB} in the limit of small $x$ only the set of diagrams
plotted in fig.~1 is important in the given gauge.
(To construct a gauge invariant amplitude one should take into account the
whole set of diagrams including the radiation from the target quarks 
\cite{GW,Baier} and also keep for the emitted and exchanged gluons the same 
mass parameter $\omega_0$.)
We take into account only the leading terms to the amplitude which are
proportional to $s$. A straightforward calculation results then in the
amplitudes corresponding to emission from the projectile quark line 
\begin{eqnarray}
M_a^{\rm rad} & = & M^{\rm el} \, 2 g \,
\frac{\vec \epsilon (\vec k_\perp - x \vec q_\perp)(1 - x)}
{(x\vec q_\perp - \vec k_\perp)^2 + (1 - x) \omega_0^2} \,
C_a,
\label{1a}
\\
M_b^{\rm rad} & = & - M^{\rm el} \, 2 g \,
\frac{\vec \epsilon \vec k_\perp (1 - x)}
{\vec k_\perp^2 + \omega_0^2} \,
C_b,
\label{1b}
\end{eqnarray}
where $C_{a,b}$ are color matrix elements associated with diagrams 
1a and 1b divided by the color factor $C^{\rm el}$ of the elastic scattering
amplitude. The elastic scattering amplitude is 
$M_{\rm el} = - 2 s q_\perp^{-2} g^2 C^{\rm el}$.
The leading term of the amplitude corresponding to the radiation 
from the gluon line via the
triple vertex is not modified, i.e. it remains the same as in case of
radiation in the vacuum,
%\begin{equation}
$M_c^{\rm rad} = M^{\rm el} 2 g 
\frac{\vec \epsilon (\vec q_\perp - \vec k_\perp) (1 - x)}
{(\vec k_\perp - \vec q_\perp)^2}
C_c,
$
%\label{1c}
%\end{equation}
where for the color factor $C_c = C_a - C_b$ holds.

In the limit of small values of $x$, where $x q_\perp \ll k_\perp$, the
sum of the amplitudes %(\ref{1a}, \ref{1b}, \ref{1c}) 
yields
\begin{equation}
M_1^{\rm rad} = M^{\rm el} \, 2 g \vec \epsilon_\perp
\left[ 
\frac{\vec k_\perp}{k_\perp^2 + \omega_0^2}
+
\frac{\vec q_\perp - \vec k_\perp}{(\vec k_\perp - \vec q_\perp)^2}
\right] C_c.
\label{eq.7}
\end{equation}
In the Abelian case $C_{a,b} = 1$, 
and $M_a^{\rm rad}$ and $M_b^{\rm rad}$ cancel.
This fact is well-known in electrodynamics: soft
photon radiation of relativistic electrons is strongly focused along the
particle's velocity, i.e. bremsstrahlung 
is suppressed as $1/s$ in the central rapidity
region where $x \to k_\perp / \sqrt{s}$. As seen in eq.~(\ref{eq.7}) this
specific non-Abelian radiation is modified by the medium polarization in the
small-$k_\perp$ region where $k_\perp \ll \omega_0$.

The r.h.s of eq.~(\ref{eq.7}) is connected with the radiation amplitude 
$R_1 = M_1^{\rm rad} / (M^{\rm el} C^{\rm el})$
which, after squaring and averaging/summing over initial/final color and
polarization states, gives the multiplicity distribution of emitted gluons
$dn^{(1)}/d^2 k_\perp d y = \overline{ | R_1 |^2}$ 
(here $y$ is the gluon rapidity)
%The resulting momentum dependence can be expressed as
\begin{equation}
\overline{ | R_1 |^2} = \frac{C_1}{\omega_0^2} \,
\frac{\omega_0^4 + 2 \omega_0^2 \vec k_\perp \vec q_\perp +
k_\perp^2 q_\perp^2}
{(\omega_0^2 + k_\perp^2)^2 \, (\vec k _\perp - \vec q_\perp)^2},
\label{eq.8}
\end{equation}
where $C_1 = 4 \pi \alpha_s \frac{N^2 -1}{N}$ for $N$ colors;
$\alpha_s = g^2/(4 \pi)$.

In the limit $k_\perp \to 0$ the spectrum (\ref{eq.8})   
is finite, i.e. the infrared divergence is removed due to the modified
dispersion relation eq.~(1). In the region
$1 \ll (\omega_0 / k_\perp)^2 \ll q_\perp / k_\perp$ the three-gluon vertex
contribution (diagram 1c)
can be neglected compared to the leading part from the 
quark-gluon vertex (diagrams 1a,b). 
We use this result later on for the analysis of the
polarization effects in the multiple scattering case.
In fig.~2, 
various contributions $\overline{ | R_1^i |^2}$ to the radiation amplitude
$\overline{| R_1 |^2}$ are displayed as a function of $k_\perp$ 
(in units of $\omega_0$) for
a given momentum transfer to show the influence of 
the modified dispersion relation. 
As can be seen in fig.~2 the polarization effect cuts off the gluon
spectrum at small values of $k_\perp$. In contrast to the vacuum case
($\omega_0 = 0$) for given $q_\perp$, in the limit $k_\perp \to 0$ 
the dominant contribution comes from the three-gluon vertex diagram.

%%%%%%%%%%%%%%%%%%%%%%%%%%%%%%%%%%%%%%%%%%%%%%%%%%%%%%%%%%%%%%%%%%%%%%

{\bf Double scattering:}
In order to consider the polarization effect in the multiple scattering
we employ the potential model \cite{GW} with with static scattering
centers. The center located at $\vec x_j$ creates a screened Coulomb
potential
$V^a(\vec q) = g T^a \exp \{ -i \vec q \vec x_j \} /(\vec q\,{}^2 + \mu^2)$,
%\begin{equation}
%V^a(\vec q) = g T^a 
%\frac{\exp \{ -i \vec q \vec x_j \} }{\vec q^2 + \mu^2},
%\label{eq.9}
%\end{equation}
where $\mu$ is the color screening mass, $T^a$ denotes a generator
of SU(N), and $\vec q$ is the spatial momentum transfer vector.
The distance between two consecutive scatterings, $\lambda$,
is assumed to be large compared to the screening length, i.e.
$\lambda \mu \gg 1$.
As mentioned above and considered in more detail in ref.~\cite{GW},
in the $A^+ = 0$ gauge
the gluon radiation off the target quarks is suppressed in the
kinematical region $\omega \gg k_\perp$, while $x$ is still small,
i.e. $x \ll 1$. Within the framework of the static scattering center model
the condition $\omega \gg k_\perp$ is associated with high rapidity of emitted
gluons. Below we focus just on the high-rapidity, soft radiation satisfying
$E \gg \omega \gg k_\perp$. It is remarkable that within such a condition the
radiation spectrum for single scattering off a static source \cite{GW}
looks the same as the spectrum eq.~(\ref{eq.8}) which is obtained  for
colliding quarks and at midrapidities.

Let us now consider two static potentials %according to eq.~(\ref{eq.9})    
separated by the distance $L$ with $L \mu \gg 1$. Similar to the case of a
single scattering there is an interval of small values of $k_\perp$
where the radiation contribution from the internal gluon lines 
according to triple gluon vertices 
and the diagram with four-gluon vertex \cite{Baier} can be neglected. 
Basing on the momentum dependence of the
corresponding amplitudes elaborated in ref.~\cite{Baier} one finds the
conditions $k_\perp \ll q_{\perp 1,2}$, 
$\vec k_\perp \ll \vec q_{\perp 1} + \vec q_{\perp 2}$, where
$\vec q_{\perp 1,2}$ denote the momentum transfers at scattering centers 1, 2.
Due to the dispersion relation (\ref{eq.1}) we obtain the additional 
restriction 
$1 \ll \left( \frac{\omega_0}{k_\perp} \right)^2 \ll \
\frac{q_{\perp 1,2}}{k_\perp}$.
In the given interval only radiation from quarks lines (cf. fig.~3)
is important. At the same time it is just the radiation from the quark lines
which is affected  by the modified dispersion relation $k^2 = \omega_0^2$.
Since we are going to demonstrate the apparency of the polarization effect for
the induced radiation we simplify the analysis and restrict ourselves to the
kinematical region where the radiation from quark lines dominates.

Under the conditions $L \mu \gg 1$ and $E \gg \omega \gg q_{\perp 1,2}$ the
factorization of the total amplitude into an elastic part and a radiation part
$R_2$ is straightforward and yields for the set of diagrams
displayed in fig.~3
\begin{equation}
R_2 = 2 g \frac{\vec \epsilon_\perp \vec k_\perp}{k_\perp^2 + \omega_0^2}
\left\{
T^{a_2} [T^{a_1}, T^b] \exp\{ikx_1\} +
[T^{a_2}, T^b] T^{a_1} \exp\{ikx_2\}
\right\} T^{a_1} T^{a_2},
\label{eq.10}
\end{equation} 
where $x_1 = (0, \vec x_1)$ and $x_2 = (t_2, \vec x_2)$ 
are the four-coordinates of two potentials with 
$t_2 = (z_2 - z_1) / v_z = L / v_z$ and $v_z$ as the longitudinal
velocity of the high-energy parton ($v_z \to 1$).
The interference between two scatterings is determined by the relative phase
factor
\begin{equation}
k (x_2 - x_1) = \omega t_2 - \vec k (\vec x_2 - \vec x_1) 
\approx 
L (\omega - k_\parallel) \equiv \frac{L}{\tau},
\end{equation}  
with the formation time $\tau = 1/(\omega - k_\parallel)$.
Due to the dispersion relation eq.~(\ref{eq.1}) the formation time is changed
from the ''vacuum'' value 
$\tau_{\rm vac} (k) = 2 \mbox{ch} y / k_\perp$ \cite{GW} to
$\tau (k, \omega_0) = 2 \mbox{ch} y / \sqrt{k_\perp^2 + \omega_0^2}$
in a medium. In the region $k_\perp \ll \omega_0$ the reduction of the
formation time due to the medium polarization can be considerable.

%%%%%%%%%%%%%%%%%%%%%%%%%%%%%%%%%%%%%%%%%%%%%%%%%%%%%%%%%%%%%%%%%%%%%%%%

{\bf Multiple scattering:}
To estimate the effect connected with the reduction of the formation time in
the medium we need to average over the interaction points $\vec x_j$ in the
general case of multiple scattering. Following the procedure developed in
ref.~\cite{GW} within the eikonal approximation one can perform the averaging
of the phase factors by
$\langle \exp \{i k (x_j - x_l) \} \rangle \approx 
\left( 1 - \frac{i \lambda}{\tau (k, \omega_0)} \right)^{l-j}$
%\begin{equation}
%< \exp \{i k (x_j - x_l) \} > \approx 
%\left( 1 - i \frac{\lambda}{\tau (k, \omega_0)} \right)^{l-j}
%\label{eq.12}
%\end{equation}
using a distribution of the length between successive scatterings,
$L_j = z_{j+1} - z_j$, $d P = \lambda^{-1} \exp \{-L_j / \lambda \} \, dL_j$.
As a result the spectrum for $m$-fold scattering can be expressed in the form
\begin{equation}
\frac{d n^{(m)}}{d^2 k_\perp \, dy} = C_m (k, \omega_0) \,
\frac{d n^{(1)}}{d^2 k_\perp \, dy},
\label{eq.13}
\end{equation} 
where the radiation spectrum for the single scattering is defined by
eq.~(\ref{eq.8}) and the radiation formation factor is for not too small
values of $m$
\begin{equation}
C_m (k, \omega_0) \approx m \frac{\chi^2}{1 + \chi^2}
+
\frac{1 -(1 - r_2) \chi^2}{r_2 (1 + \chi^2)^2}
\label{eq.14}
\end{equation}
with $ \chi (k, \omega_0) = \lambda / ( r_2 \tau (k, \omega_0))$
and $r_2 = C_A /(2 C_2)$;
$C_2 = C_A = N$ for gluons,
$C_2 = C_F = (N^2 - 1)/(2N)$ for quarks.
It should be stressed that, while the dependence of the radiation formation
factor $C_m$ on the parameter $\chi$ looks the same as in the vacuum case
\cite{GW}, this parameter itself is obviously modified by the reduction of the
formation time in the medium. In fig.~4 the dependence of $C_m$ on
$k_\perp$ is displayed for $m = 10$ for various values of the emitted gluon
energy which defines the dimensionless parameter 
$a_0 = \lambda \omega_0^2 / (2 r_2 \omega)$.
As seen in fig.~4, due to the medium reduction of the formation time
the radiation formation factor $C_m$,
related to the LPM effect, is also modified,
in particular in the region $k_\perp < \omega_0$.

%%%%%%%%%%%%%%%%%%%%%%%%%%%%%%%%%%%%%%%%%%%%%%%%%%%%%%%%%%%%%%%%%

{\bf Summary:}
In summary we have demonstrated the existence of a QCD analogue of the 
Ter-Mikaelian effect. In doing so a certain phase space
region is selected where the treatment simplifies. The effect consists
in the suppression of the induced radiation and the modification of the 
spectrum at low $k_\perp$ due to the refractive
properties of the medium.

{\bf Acknowledgments:}
Useful discussions with P. Levai,
W. M\"uller, A.M. Snigirev, B.G. Zakharov,
and G.M. Zinovjev
are gratefully acknowledged.
The work is supported by the grants BMBF 06DR928/1, WTZ UKR-008-98
and STCU 015.

%%%%%%%%%%%%%%%%%%%%%%%%%%%%%%%%%%%%%%%%%%%%%%%%%%%%%%%%%%%%%%%% 
\newpage
%{\small

%}
%%%%%%%%%%%%%%%%%%%%%%%%%%%%%%%%%%%%%%%%%%%%%%%%%%%%%%%%%%%%%%%%%%%%%%%

\begin{figure}[h]
{\epsfig{file=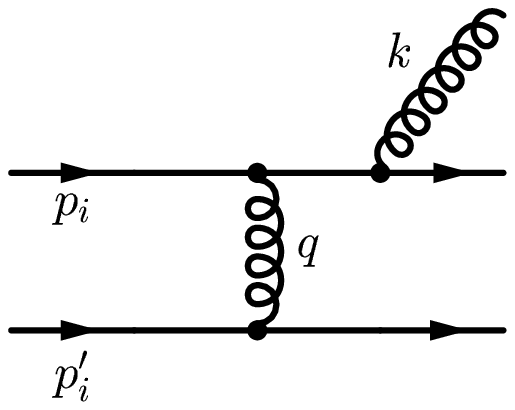,width=4.5cm}}
\hspace*{\fill}
{\psfig{file=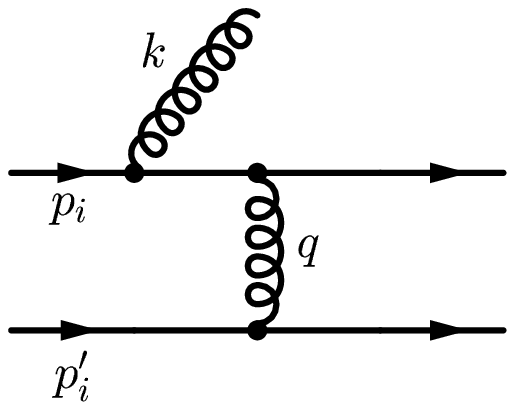,width=4.5cm}}
\hspace*{\fill}
{\psfig{file=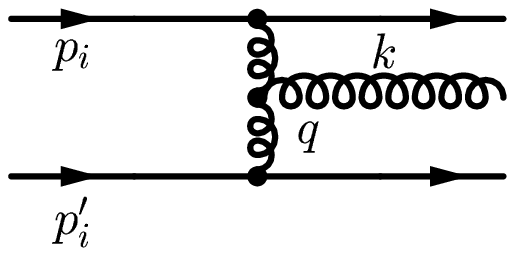,width=4.5cm}}
\caption{
Feynman diagrams for single scattering 
which give the leading contribution to one-gluon emission in the 
$A^+ = 0$ gauge.
The upper (lower) line is called projectile (target) quark line.
}
\label{fig.1}
\end{figure}

\begin{figure}[h]
\centerline{{\psfig{file=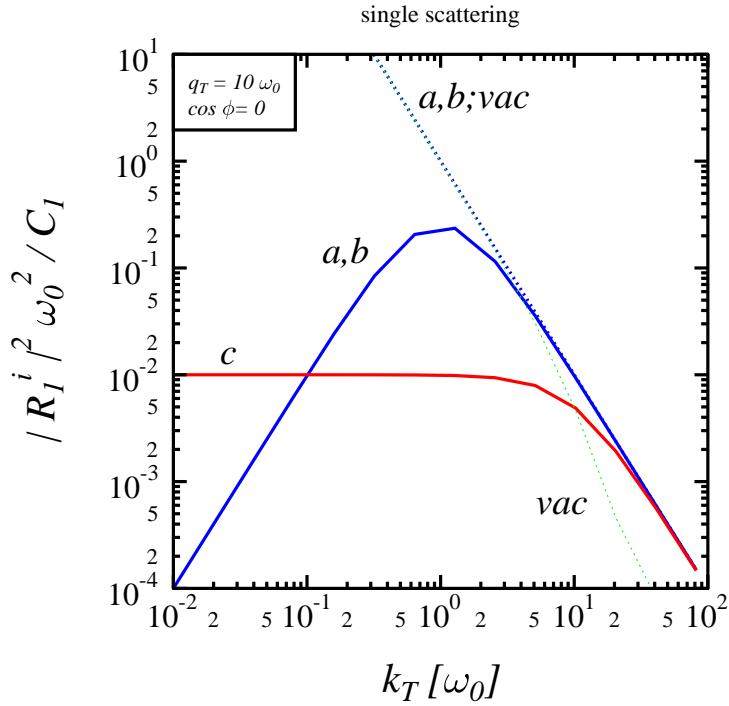,width=10cm,angle=0}}}
%~\\[3mm]
\caption{
Various contributions $|R_1^i|^2$ to the radiation amplitude for single scattering
as a function of $k_\perp$
(in units of $\omega_0$) for $q_\perp = 10 \omega_0$.
''a,b'' (''a,b; vac'') is for the radiation contribution from the quark line
for $\omega_0 \ne 0$ (for vacuum, i.e. $\omega_0 = 0$),
''c'' is the contribution from the three-gluon vertex diagram,
and ''vac'' labels the total amplitude in vacuum.
To avoid complications with the collinear singularity at
$\vec k_\perp \parallel \vec q_\perp$ we have chosen
$\vec k_\perp \perp \vec q_\perp$. 
}
\label{fig.2}
\end{figure}

\begin{figure}[h]
{\psfig{file=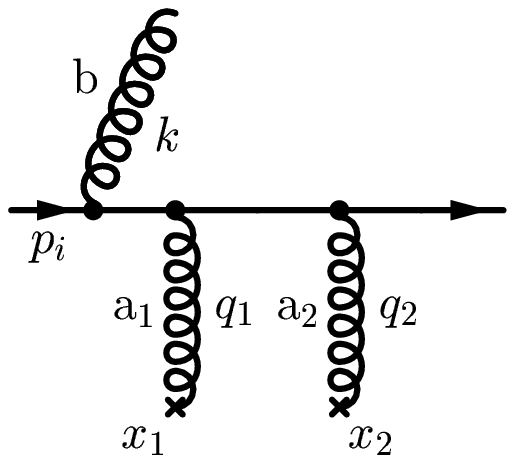,width=4.5cm}}
\hspace*{\fill}
{\psfig{file=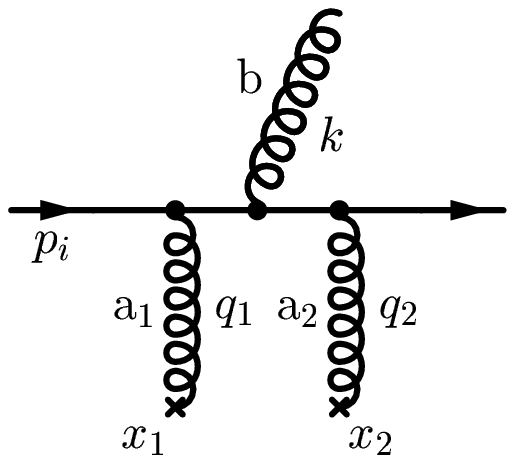,width=4.5cm}}
\hspace*{\fill}
{\psfig{file=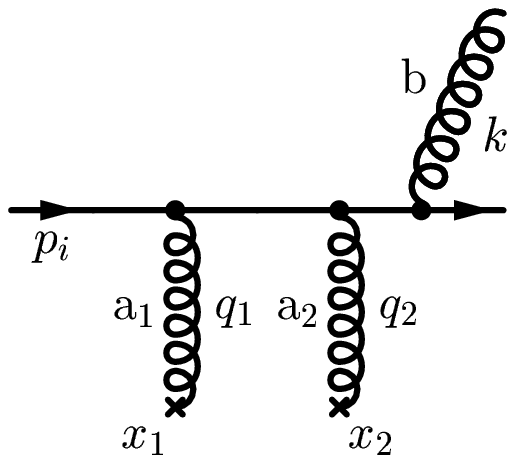,width=4.5cm}}
\caption{
Relevant Feynman diagrams for double scattering at static centers.
}
\label{fig.3}
\end{figure}

\begin{figure}[h]
\centerline{{\psfig{file=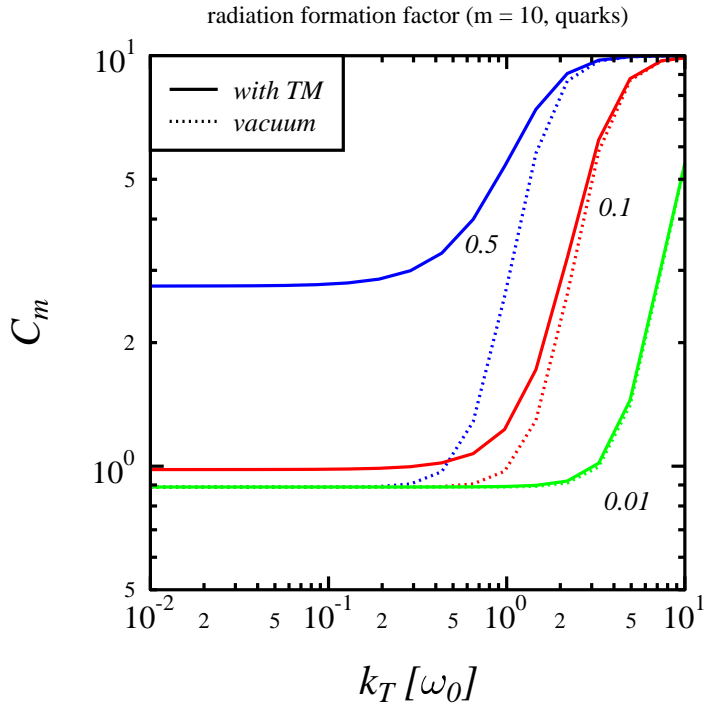,width=10cm,angle=0}}}
\caption{
The radiation formation factor for quarks 
as a function of $k_\perp$ (in units of $\omega_0$) 
for $m = 10$.
The curves are labeled by the value of $a_0$;
full (dotted) curves: with (without) Ter-Mikaelian effect.
}
\label{fig.4}
\end{figure}

\end{document}